\documentclass[manuscript]{aastex}

\usepackage{natbib}
\usepackage{aas_macros}
\usepackage{color}
\usepackage{amsmath}
\usepackage{bm}
\newcommand{\myemail}{h.iijima@eps.s.u-tokyo.ac.jp}

\definecolor{gray}{gray}{0.7}

\newcommand{\figdir}{./}

\shorttitle{Effect of coronal temperature}
\shortauthors{Iijima and Yokoyama}

\begin{document}


\title{
Effect of coronal temperature \\
on the scale of solar chromospheric jets
}


\author{H. Iijima and T. Yokoyama}
\affil{
Department of Earth and Planetary Science, The University of Tokyo,
7-3-1 Hongo, Bunkyo-ku, Tokyo 113-0033, Japan.
}
% \author{H. Iijima\altaffilmark{1} and T. Yokoyama\altaffilmark{1}}
% \affil{
% Department of Earth and Planetary Science, University of Tokyo.
% }
\email{\myemail}


% \altaffiltext{1}{
% Department of Earth and Planetary Science, University of Tokyo,
% 7-3-1 Hongo, Bunkyo-ku, Tokyo 113-0033, Japan.
% }


\begin{abstract}
 We investigate the effect of coronal temperature
 on the formation process of solar chromospheric jets
 using two-dimensional magnetohydrodynamic simulations
 of the region from the upper convection zone
 to the lower corona.
 We develop a new radiative magnetohydrodynamic code
 for the dynamic modeling of the solar atmosphere,
 employing a LTE equation of state,
 optically thick radiative loss in the photosphere,
 optically thin radiative loss in the chromosphere and the corona,
 and thermal conduction along the magnetic field lines.
 Many chromospheric jets are produced in the simulations
 by shock waves passing through the transition region.
 We find that these jets are projected farther outward
 when the coronal temperature is lower
 (similar to that in coronal holes)
 and shorter when the coronal temperature is higher
 (similar to that in active regions).
 When the coronal temperature is high,
 the deceleration of the chromospheric jets
 is consistent with the model
 in which deceleration is determined
 by the periodic chromospheric shock waves.
 However, when the coronal temperature is low,
 the gravitational deceleration becomes more important
 and the chromospheric jets approach ballistic motion.
\end{abstract}





%%%%%%%%%%%%%%%%%%%%%%%%%%%%%%%%%%%%%%%%%%%%%%%%%%%%%%%%%%%%%%%%%%%%%%
\section{Introduction}\label{sec_introduction}
%%%%%%%%%%%%%%%%%%%%%%%%%%%%%%%%%%%%%%%%%%%%%%%%%%%%%%%%%%%%%%%%%%%%%%

% \subsection{背景: 彩層ジェットの領域依存性}

Various chromospheric jets are observed
in the solar atmosphere.
In quiet regions and coronal holes,
we observe classical (type I) spicules at the limb,
which have maximum lengths of 4--10 Mm,
lifetimes of 1--7 min,
and maximum upward velocities of 20--100 km/s
\citep{1972ARA&A..10...73B}.
Recently, the existence of more violent
``type II'' spicules is reported
\citep{2007PASJ...59S.655D}.
The type I and II spicules are considered to
have their on-disk counterparts of
mottles \citep{2007ApJ...660L.169R}
and the rapid blue-shifted events (RBEs)
\citep{2009ApJ...701L...1D,2009ApJ...705..272R,2009ApJ...707..524M},
respectively.
The dynamic fibrils observed on the disk in active regions
have maximum lengths of several Mm,
lifetimes of 2--8 min,
and maximum velocities of 10--40 km/s
\citep{2006ApJ...647L..73H,2007ApJ...655..624D}.
\cite{2007ApJ...660L.169R} reported that
the dynamic fibrils are considered to be
the active region counterpart of the quiet-sun mottles.

% \subsection{背景: 衝撃波駆動ジェットの理論研究}

Theoretical studies have suggested
various models for these chromospheric jets
\citep[see][for a review]{2000SoPh..196...79S}.
For the dynamic fibrils,
the periodic acoustic wave model is in agreement with observations
\citep{2006ApJ...647L..73H,2007ApJ...655..624D,2007ApJ...666.1277H},
particularly in terms of the correlation
between deceleration and maximum velocity.
The origin of spicules (types I and II) is still under debate.
Many candidates, such as
acoustic waves \citep[e.g.,][]{1982ApJ...257..345H,1982SoPh...75...99S},
Alfv\'en waves \citep[e.g.,][]{1982SoPh...75...35H,1999ApJ...514..493K},
and magnetic reconnection
\citep[e.g.,][]{1969PASJ...21..128U,2009ApJ...702....1H,2013PASJ...65...62T}
have been suggested.
Most of these models include a process by which
the transition region is lifted by the shock wave
propagating upward into the corona.
This process is called the shock-transition region interaction
\citep{1982ApJ...257..345H}.
Several multi-dimensional simulations with
the sophisticated modeling of convective motion have been reported
\citep{2006ApJ...647L..73H,2009ApJ...701.1569M,2011ApJ...743..142H}.
Although some of these studies is intended
to be the representative for quiet regions,
the simulated chromospheric jets was significantly shorter
than the observed spicules.
% Although these simulations succeed in reproducing
% the relatively small-scale chromospheric jets
% similar to the active region dynamic fibrils,
% the larger-scale features similar to
% the coronal hole spicules are not reproduced.

% \subsection{目的}

Using idealized one-dimensional hydrodynamic simulations,
\cite{1982SoPh...78..333S} explained
why the spicules are long in the coronal hole.
The key process is the amplification
of the strength of the shock wave in the solar chromosphere.
Based on their idea,
we investigate the effect of the coronal temperature
on the scale of the solar chromospheric jets
using two-dimensional magnetohydrodynamic simulations
with more realistic physical processes,
to provide a unified perspective
on the quiet region spicules, the coronal hole spicules,
and the active region dynamic fibrils.

%%%%%%%%%%%%%%%%%%%%%%%%%%%%%%%%%%%%%%%%%%%%%%%%%%%%%%%%%%%%%%%%%%%%%%
\section{Numerical Model}\label{sec_method}
%%%%%%%%%%%%%%%%%%%%%%%%%%%%%%%%%%%%%%%%%%%%%%%%%%%%%%%%%%%%%%%%%%%%%%

% \subsection{基礎方程式}

We develop a new radiation magnetohydrodynamic code
for the dynamical modeling of the solar atmosphere.
The equations solved by this code include gravity,
a LTE equation of state,
the radiative cooling,
and thermal conduction along the magnetic field line.

% \subsubsection{磁気流体方程式}

The solar atmosphere is filled with strong shock waves
with a wide range of plasma beta.
A robust numerical scheme is required
for the simulation of these regions.
We employ a higher-order magnetohydrodynamic scheme
based on the constrained transport method \citep{1988ApJ...332..659E}
with divergence-free magnetic field reconstruction
\citep{2009JCoPh.228.5040B}.
% \tgray{
% The fifth-order WENO-Z reconstruction is employed
% with the third-order optimal strong stability preserving (SSP)
% Runge--Kutta method \citep{1996JCoPh.126..202J}.
% The local Lax--Friedrich scheme is employed as the Riemann solver.
% }
The equations are solved in the conservative form.
We do not consider explicit viscosity or magnetic diffusivity.

% \subsubsection{状態方程式}

We assume the local thermodynamic equilibrium for the equation of state.
The six most abundant elements are accounted
with the metal mass fraction of \cite{2006CoAst.147...76A}.
The formation of molecular hydrogen is not considered.
% \tgray{
% The six most abundant elements are
% accounted for the equation of state,
% assuming local thermodynamic equilibrium.
% We assume the same metal mass fraction as \cite{2006CoAst.147...76A}.
% Two bound levels and the ionized state are considered
% for the hydrogen state using the Saha--Boltzmann equations.
% The formation of molecular hydrogen is not considered.
% The other elements are assumed to be in Saha's ionization equilibrium
% with the partition function from \cite{1981ApJS...45..621I}.
% The equation of state is pre-calculated,
% stored in the numerical table,
% and interpolated during time-integration.
% }

% \subsubsection{輻射加熱項}

The radiative cooling term is evaluated as
a combination of the two different approximations.
In the photosphere and lower chromosphere,
the radiative transfer equation is solved directly
for the better modeling of the convective motion
\citep{1982A&A...107....1N}.
The short characteristic method \citep{1988JQSRT..39...67K}
is employed with the A4 quadrature in \cite{carlson1963methods}.
The gray approximation is assumed
with the OPAL Rosseland mean opacity
\citep{1996ApJ...464..943I} in the high-temperature region
and the opacity of \cite{2005ApJ...623..585F}
in the low-temperature region.
In the upper chromosphere and corona,
the optically thin radiative loss function is employed
to estimate the radiative cooling.
The loss function is calculated from
the CHIANTI atomic database \citep{2012ApJ...744...99L},
with the extension by \cite{2012ApJ...751...75G}
for the chromospheric temperature.
These photospheric and chromospheric cooling terms
are switched by the function of the column mass density.
In addition to these terms,
we include an artificial heating term
which becomes active only
when the temperature falls below 2,500 K.
% In addition to these physical cooling terms,
% an artificial heating term is included
% to prevent the temperature from being lower than 2,500 K.
% \tgray{
% In addition to these physical cooling terms,
% an artificial heating term is included
% to prevent very low temperatures around
% the photosphere and low chromosphere.
% This artificial heating term is switched on
% only when the temperature becomes lower than 2,500 K
% and heats the plasma over a timescale of 1 s.
% }

% \subsubsection{非等方熱伝導}

Spitzer-type anisotropic thermal conduction is treated by
second-order operator splitting and
the second-order super-time-stepping method \citep{2012MNRAS.422.2102M}.
The flux-limiting method \citep{2007JCoPh.227..123S} is employed
to maintain the monotonicity of the temperature.

% \subsection{初期条件、境界条件、グリッドサイズ}

% \subsubsection{領域サイズ、グリッドサイズ}

The two-dimensional numerical domain
in the $XZ$-plane extends
from 2 Mm below the surface to 14 Mm above.
The horizontal extent is 18 Mm.
The uniform grid size is 42 km in the horizontal direction
and 32 km in the vertical direction.

% \subsubsection{境界条件}

The bottom boundary condition is the ``open'' boundary.
The total (gas plus magnetic) pressure equilibrium is assumed.
The entropy of the upward plasma is fixed.
The upward velocity field is
slowly damped to the horizontally uniform upflow
which complements the mass flux by the downward flow
\cite[e.g.,][]{2002PhDT........16B}.
This weak damping is not important in the present results
but able to suppress the unrealistic instability
which appears after the long time integration.
The downward plasma evolves adiabatically and freely.
The top boundary has a free-slip condition
and is open for vertical flow.
The density is extrapolated exponentially
by the scale height at the boundary.
Our two-dimensional simulation
with a top boundary at the lower corona
cannot maintain the 1 MK coronal temperature
\citep[e.g.,][]{2011A&A...530A.124L,2011ApJ...743..142H}.
To prevent the corona from being cooled
by radiation and thermal conduction,
the conductive flux is introduced through
the top boundary to heat the corona.
In this study, the conductive flux is adjusted
to preserve the different coronal temperature
at the top boundary, $T_\mathrm{c}$.
The magnetic field is assumed to be vertical
at the top and bottom boundaries.
The periodic boundary condition is employed horizontally.

% \subsubsection{初期条件}

The initial condition is the plane-parallel atmosphere
with a uniform vertical magnetic field of 3 G.
The sufficiently relaxed convection is obtained
after the integration of 5 solar hours
with a temperature at the top boundary,
$T_\mathrm{c}$, of 1 MK.
We impose the uniform vertical magnetic field of 30 G
on the relaxed atmosphere
and integrate another 3 solar hours with the same $T_\mathrm{c}$.
Next, we change the temperature at the top boundary,
$T_\mathrm{c}$, to 2, 1, and 0.4 MK
to imitate the conditions of the active region,
the quiet region, and the coronal hole, respectively.
The simulations are integrated for another 1 solar hour
after the change of $T_\mathrm{c}$.
We find that the coronal density and temperature
reach statistical equilibrium within 20 min.
The data analyzed below is the last 30 min of each simulation.

%%%%%%%%%%%%%%%%%%%%%%%%%%%%%%%%%%%%%%%%%%%%%%%%%%%%%%%%%%%%%%%%%%%%%%
% \section{Results and Discussion}\label{sec_results}
\section{Results}\label{sec_results}
%%%%%%%%%%%%%%%%%%%%%%%%%%%%%%%%%%%%%%%%%%%%%%%%%%%%%%%%%%%%%%%%%%%%%%

% \subsection{第一感の結果}

\begin{figure}
 % \epsscale{2.2}
 \plotone{\figdir/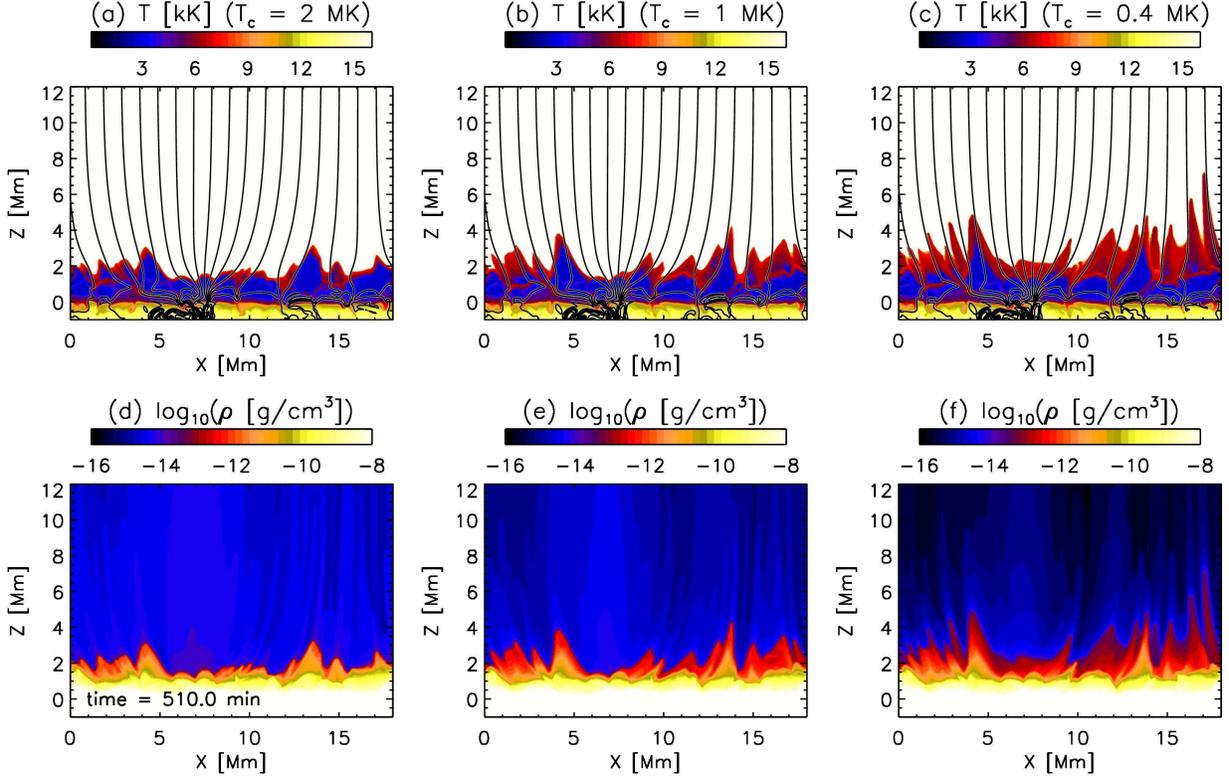}
 \caption{
 Snapshots from the three simulations
 with the different temperatures at the top boundary, $T_\mathrm{c}$.
 Panels (a)--(c) show the temperature,
 and panels (d)--(f) represent the logarithm of mass density.
 The solid lines in panels (a)--(c) indicate the magnetic field lines.
 Panels (a) and (d) correspond to the simulation
 with $T_\mathrm{c}=2$ MK,
 panels (b) and (e) correspond to the simulation
 with $T_\mathrm{c}=1$ MK,
 and panels (c) and (f) correspond to the simulation
 with $T_\mathrm{c}=0.4$ MK.
 (An animation of this figure is available
 in the electronic edition of {\it The Astrophysical Journal Letters}.)
 }
 \label{fig01}
\end{figure}

Figure \ref{fig01} shows snapshots from three simulations
with the temperature at the top boundary,
$T_\mathrm{c}$, set to 2, 1, and 0.4 MK.
We find the strong magnetic field concentration near $X=8$ Mm.
Because the magnetic field structure is essentially the same
among these three simulations,
we can concentrate on the effect of the coronal temperature.
The jet-like structures with various scales are formed in the chromosphere.
These structures are basically driven
by the shock waves, arising from convective motion,
which pass through the transition region.
This process is consistent with
both the one-dimensional simulations
\citep{1982SoPh...75...35H,1982SoPh...75...99S}
and multi-dimensional simulations \citep{2011ApJ...743..142H}.
We clearly find that the simulation
with a lower coronal temperature
produces higher chromospheric jets.
The density near the top of each jet
and over the entire corona is smaller when $T_\mathrm{c}$ is lower.
The widths of the produced jets in these simulations are
approximately independent of $T_\mathrm{c}$ if measured
at a specific height like $Z=3$ Mm,
but the jets become sharper near their tops
in the lower $T_\mathrm{c}$ simulations.

% \subsection{統計的な比較}

We automatically detect the chromospheric jets in the simulations
for the purpose of quantitative comparison.
The procedure is basically same as that in \cite{2011ApJ...743..142H}.
The vertical motion of the transition region
along the magnetic field line
is fitted by a parabola.
The maximum length
(i.e., the maximum height of transition region
evaluated from the initial height),
the lifetime, the maximum upward velocity,
and the deceleration for each jet
are calculated from the fitting.
The result of the statistical analysis is
shown in Figure \ref{fig02}.
We find a clear correlation between
maximum length and maximum upward velocity
for all three simulations (panels (a)--(c)).
The lifetime roughly correlates with the maximum velocity
for each simulation (panels (d)--(f)).
The lifetime becomes longer and the correlation between
lifetime and maximum velocity becomes stronger
when the coronal temperature is lower.
The correlation between deceleration
and maximum velocity is relatively weak (panels (g)--(i)).
The deceleration for $T_\mathrm{c}=2$ MK (panel (g))
is distributed over a relatively broad range,
exceeding the gravitational deceleration on the solar surface.
The deceleration for $T_\mathrm{c}=0.4$ MK (panel (i))
is concentrated below the gravitational deceleration.

The abovementioned dependence
on the temperature at the top boundary, $T_\mathrm{c}$,
can be explained as follows.
When $T_\mathrm{c}$ is high (i.e., 2 MK)
and the chromospheric jets are small,
as in panels (a), (d), and (g) in Figure \ref{fig02},
the obtained correlations are consistent with
the ``shock deceleration hypothesis''
suggested by \cite{2007ApJ...666.1277H}.
According to this hypothesis,
the deceleration is determined
by the periodicity of the slow-mode shock wave
and can exceed the gravitational deceleration
\citep[see Eq. (1) in][]{2007ApJ...666.1277H}.
The lifetime of the jet is determined
by the period of the driving shock wave.
The dotted lines in panels (a), (d), and (g)
are the theoretical lines of parabolic motion
with lifetimes of 2, 3, and 5 min.
These lines are well-consistent with the results
from the simulation with $T_\mathrm{c}=2$ MK.
Because the lifetime of chromospheric jet is
equal to the period of driving acoustic wave
in the ``shock deceleration hypothesis'',
the long lifetime in $T_\mathrm{c}=0.4$ MK case (panels (f))
requires long period of driving waves, which is not expected
under the short acoustic cutoff period in the lower solar atmosphere
\citep[e.g.][]{1973SoPh...30...47M}.
The deceleration by the gravity becomes
more important in this case.
The resulting motion of taller chromospheric jets approaches free-fall.
The dashed lines in panels (c), (f), and (i) indicate
the theoretical lines of ballistic motion,
which approximately explain the properties of simulated jets.
Since the simulated jets in the case of $T_c=0.4$ MK
do not exactly follow ballistic paths,
both of the gravitational and gas pressure gradient forces play a role
in addition to the inclination of the magnetic field line.
% \tred{
% We note that the simulated jets in $T_c=0.4$ MK case
% does not follow the ballistic path exactly
% and the gas pressure gradient force
% and the inclination of the magnetic field play a role
% as shown in the next paragraph.
% }
% \tred{
% We note that the relation in panel (c) can also be approximated
% by the theoretical lines of parabolic motion
% with lifetimes of 2 and 7 min.
% However, this long lifetime of 7 min is difficult to be explained
% by the ``shock deceleration hypothesis''
% considering the short acoustic cutoff period as discussed above.
% }
The case of $T_\mathrm{c}=1$ MK, shown in panels (b), (e), and (h),
exhibits intermediate properties between these two extremes.

% \subsection{減速率決定のメカニズム}

To consider the abovementioned view
on the determination of deceleration,
we investigate the vertical force
near the top of the chromospheric jets.
The vertical component of the equation of motion
can be written as follows.
\begin{align}
 \begin{split}
  -\frac{DV_z}{Dt}
  &=D^G_\parallel+D^P_\parallel+D^G_\perp+D^P_\perp+D^L_\perp
  \\
  D^G_\parallel
  &=g_0b_z^2
  \\
  D^P_\parallel
  &=\bm{b}\cdot\left(\frac{1}{\rho}\nabla P\right)b_z
  \\
  D^G_\perp
  &=g_0-D^G_\parallel=g_0b_x^2
  \\
  D^P_\perp
  &=\frac{1}{\rho}\frac{\partial P}{\partial z}-D^P_\parallel
  \\
  D^L_\perp
  &=-\frac{1}{\rho}\left(\bm{J}\times\bm{B}\right)_z
 \end{split}
\end{align}
Here, $g_0$ is the gravitational acceleration at the solar surface,
$P$ is the gas pressure,
$\bm{B}$ is the magnetic flux density,
$\bm{J}$ is the current density,
$D/Dt=\partial/\partial t+\bm{V}\cdot\nabla$
is the Lagrangian derivative,
and $\bm{b}=\bm{B}/\left|\bm{B}\right|$
is a unit vector parallel to the magnetic field.
If we assume that the transition region
at the top of the chromospheric jets
is the contact discontinuity,
the deceleration of the jets
can be approximated by $-DV_z/Dt$
near the transition region.
The result is shown in Figure \ref{fig03}.
% The terms of the equation of motion
% is evaluated at the height of $200$--$600$ km
% below the transition region
% and averaged over the lifetime of each jet.
% The data within 25 s before and after
% the initial/final time of the chromospheric jets
% are removed from the analysis
% to avoid the effect of the shock front
% passing through the transition region.
Panels (a)--(c) show the deceleration
produced by the total force parallel to the magnetic field
$D^P_\parallel+D^G_\parallel$
versus the deceleration calculated
by the motion of the chromospheric jets.
Although there is some level of dispersion,
the motion of the transition region
is roughly explained by the forces
parallel to the magnetic field.
The possible candidates of the discrepancy
are the effect of the forces
perpendicular to the magnetic field
($D^G_\perp$, $D^P_\perp$, and $D^L_\perp$),
the deviation of the transition region
from the contact discontinuity
by the thermal conduction and waves,
and the fitting and discretization errors.
Panels (d)--(f) show the deceleration produced by
the gravity parallel to the magnetic field $D^G_\parallel$.
$D^G_\parallel$ is more concentrated
near the value of gravitational acceleration
in the lower coronal temperature.
Because the structure of the magnetic field
is nearly independent of the coronal temperature,
the temperature dependence in these panels
are caused by the difference of the height of produced jets.
When the coronal temperature is higher,
the maximum length of jets becomes small
and the transition region moves
near the root of the magnetic field
where the magnetic field is more inclined.
% We also find moderate correlation
% between $D^G_\parallel$ and the deceleration of jets.
% This is interpreted that the jets with small deceleration
% これは減速率が小さいジェットの多くが最大速度が小さく
% (Figure 2, panels (g)--(i))、
% ジェットの背も低いためであると解釈出来る。
Panels (g)--(i) show the deceleration produced by
the gas pressure gradient force
parallel to the magnetic field $D^P_\parallel$.
% The pressure gradient force basically affects
% opposite to the direction of gravity
% and contributes to decrease the deceleration of jets.
In $T_c=0.4$ MK case,
the pressure gradient force always
decreases the deceleration of the jets.
However, in $T_c=2$ and $1$ MK cases,
a part of the jets is decelerated
by the pressure gradient force.
This is consistent with the ``shock deceleration hypothesis''
where the deceleration is determined
by the periodic shock wave.

% \subsection{差異が出る理由}

To understand
why the scale of the chromospheric jets
depends on the coronal temperature,
we investigate the relationship between the coronal gas pressure
and the maximum velocity of the produced jets
as shown in Figure \ref{fig04}.
The scale height of the gas pressure in the corona
is sufficiently long and
the gas pressure is almost constant across the transition region.
Thus, we can use the coronal gas pressure, $P_\mathrm{c}$,
as the representative value of the gas pressure near the transition region.
\cite{1982SoPh...78..333S} suggest that
the density difference at the transition region
causes the differences
in the amplitude of the chromospheric shock wave
and in the size of the chromospheric jets.
For a linear wave, the energy flux
of the slow magneto-acoustic wave is conserved during the propagation
as $S\rho V^2C_s=\mathrm{constant}$.
Here $S$ is the cross-section of the magnetic flux tube,
$C_s$ is the speed of the slow magneto-acoustic wave,
and $V$ is the velocity parallel to the magnetic field.
This relation is reduced to
$V\propto (S\rho C_s)^{-1/2}\sim (SP_\mathrm{c})^{-1/2}$.
The dependence on $S$ does not appear in our current results
because the magnetic field structure
is almost the same in each of our three simulations.
In fact, the shock wave is strong
in the upper chromosphere,
and the energy dissipation near the shock front
violates the energy flux conservation.
\cite{1960PThPh..23..294O} analytically derived
the relation of $V\propto P_\mathrm{c}^{-0.236}$
for a strong shock wave with
the constant chromospheric temperature
and an adiabatic heat ratio of $5/3$.
\cite{1982SoPh...78..333S} reported
the relation of $V\propto P_\mathrm{c}^{-0.23}$
in their one-dimensional adiabatic hydrodynamic simulation,
which is very close to the analytic estimate for strong shock wave.
We obtain the relation of
$V\propto P_\mathrm{c}^{-0.089}$
from the three simulations with different coronal temperatures.
This result indicates that another energy-loss process
contributes to the formation of chromospheric jets in our simulations.
The possible candidates are the radiative energy loss,
the energy leakage by multidimensional effects,
the thermal conduction,
and the latent heat of ionization.
The shock-transition region interaction \citep{1982ApJ...257..345H}
and the dynamic variation of the chromospheric temperature
will also affect the result.
We note that the amplitude of the maximum upward velocity depends
only on the nonlinear propagation of acoustic waves
in the photosphere and chromosphere
and does not depend on the deceleration models
discussed in the preceding paragraphs.
% The coronal gas pressure dependence
% of the maximum length is affected by the value of the deceleration.

%%%%%%%%%%%%%%%%%%%%%%%%%%%%%%%%%%%%%%%%%%%%%%%%%%%%%%%%%%%%%%%%%%%%%%
\section{Discussion}\label{sec_discussion}
%%%%%%%%%%%%%%%%%%%%%%%%%%%%%%%%%%%%%%%%%%%%%%%%%%%%%%%%%%%%%%%%%%%%%%

% \subsection{観測研究との関係}

\begin{figure}
 % \epsscale{2.2}
 \plotone{\figdir/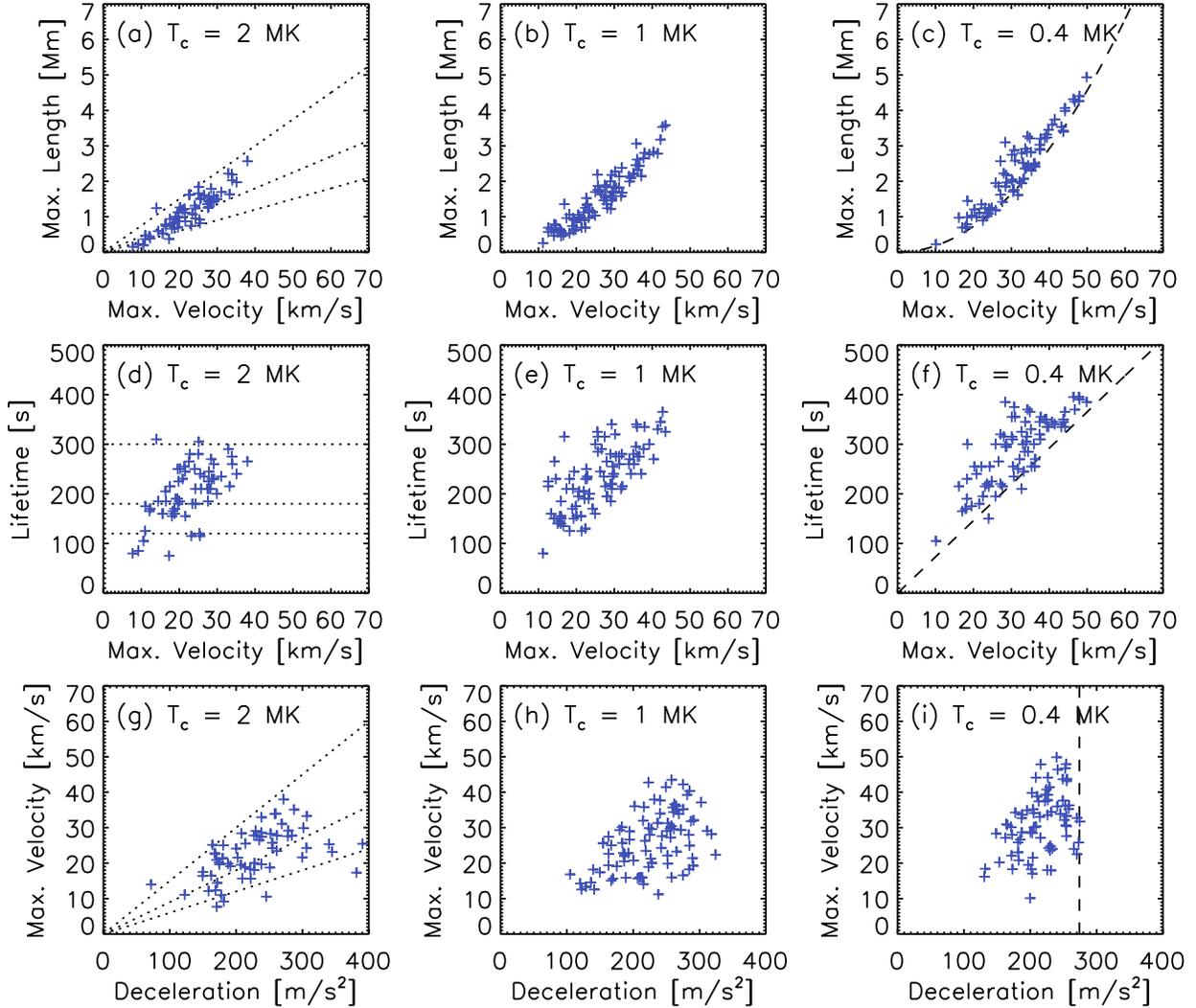}
 \caption{
 Statistical analysis of the produced chromospheric jets
 in the three simulations with different temperatures
 at the top boundary, $T_\mathrm{c}$.
 Each cross sign corresponds to a chromospheric jet detected.
 See Section \ref{sec_results} for the description
 on the procedure for chromospheric jet detection.
 Panels (a), (d), and (g) correspond to $T_\mathrm{c}=2$ MK,
 panels (b), (e), and (h) correspond to $T_\mathrm{c}=1$ MK,
 and panels (c), (f), and (i) correspond to $T_\mathrm{c}=0.4$ MK.
 The three dotted lines in panels (a), (d), and (g)
 represent the theoretical lines
 of parabolic motion with lifetimes of 2, 3, and 5 min.
 The dashed lines in panels (c), (f), and (i)
 indicate the theoretical lines of ballistic motion.
 }
 \label{fig02}
\end{figure}

\begin{figure}
 % \epsscale{2.2}
 \plotone{\figdir/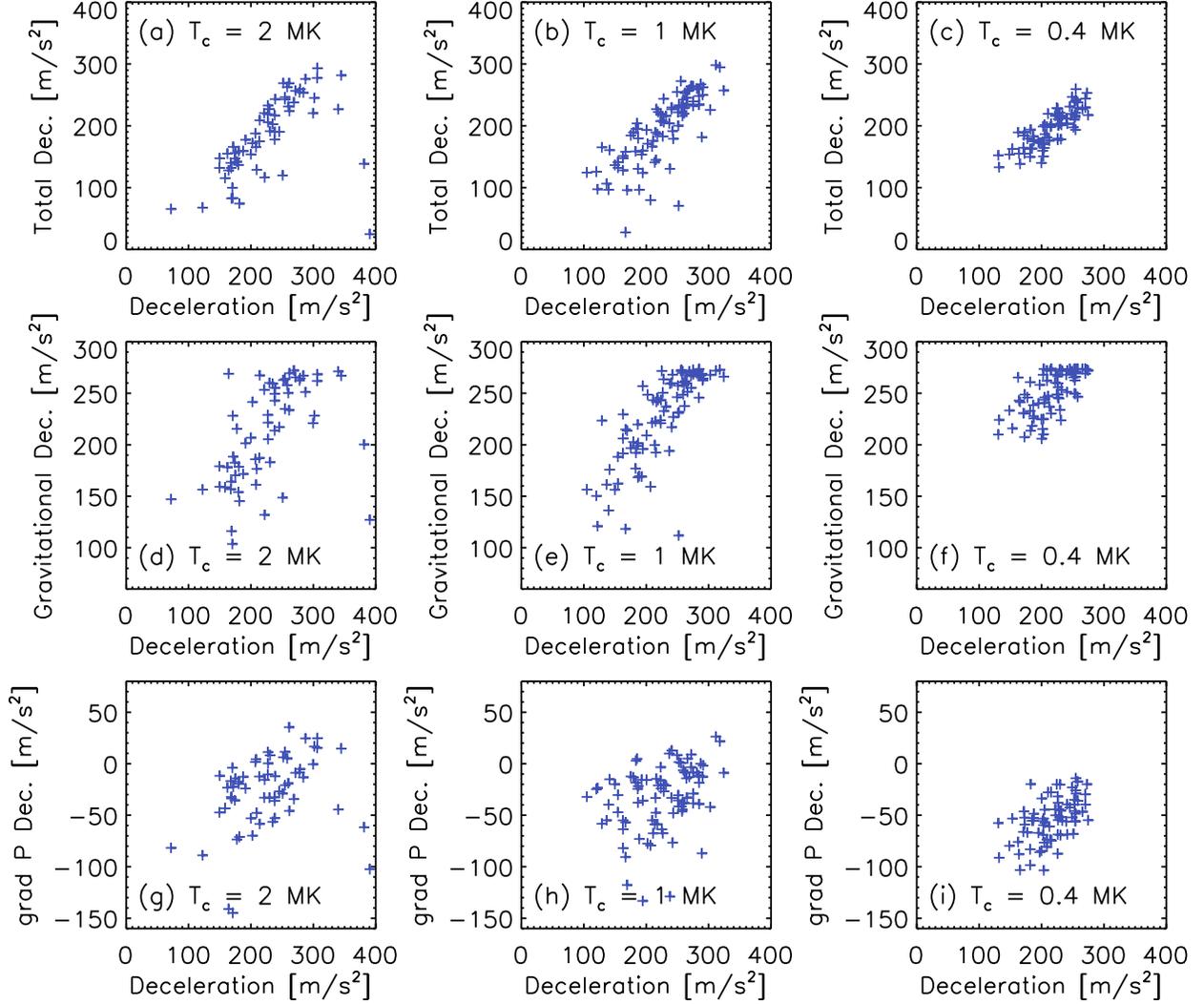}
 \caption{
 Contribution of gas pressure gradient and gravitational forces
 on the deceleration of the produced chromospheric jets
 in the three simulations with different temperatures
 at the top boundary, $T_\mathrm{c}$.
 Each cross sign corresponds to a chromospheric jet detected.
 Panels (a), (d), and (g) correspond to $T_\mathrm{c}=2$ MK,
 panels (b), (e), and (h) correspond to $T_\mathrm{c}=1$ MK,
 and panels (c), (f), and (i) correspond to $T_\mathrm{c}=0.4$ MK.
 See Section \ref{sec_results} for the description
 on the definition of each variables.
 }
 \label{fig03}
\end{figure}

\begin{figure}
 % \epsscale{2.2}
 \plotone{\figdir/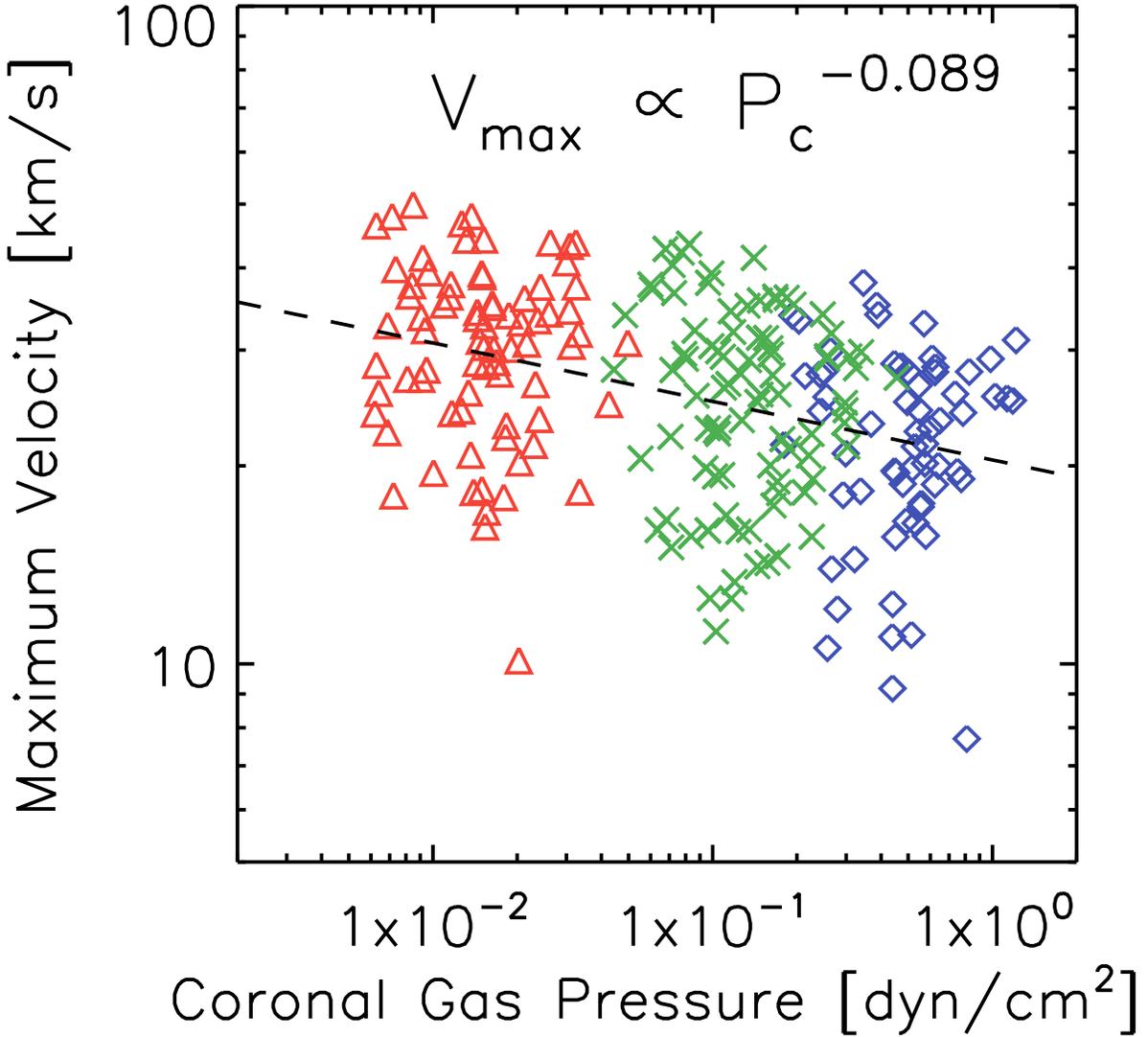}
 \caption{
 Dependence of the maximum upward velocity
 of the chromospheric jets
 on the coronal gas pressure
 from the simulation of
 $T_\mathrm{c}=2$ MK (shown as blue diamonds),
 $T_\mathrm{c}=1$ MK (shown as green crosses),
 and $T_\mathrm{c}=0.4$ MK (shown as red triangle).
 The coronal gas pressure, $P_\mathrm{c}$,
 is measured at the initial time of each chromospheric jet
 and averaged over the height range
 of 1--3 Mm above the transition region.
 The dashed line is the result of the least-square fitting,
 $V_\mathrm{max}\propto P_\mathrm{c}^{-0.089}$,
 assuming a power-law relation.
 }
 \label{fig04}
\end{figure}

We can compare our three simulations
with different temperatures at the top boundary
($T_\mathrm{c}=2,1,$ and $0.4$ MK)
to the observations of the active region dynamic fibrils,
the quiet region spicules, and the coronal hole spicules, respectively.
Panels (a), (d), and (g) in Figure \ref{fig02}
show behavior similar to the observation
of the active region dynamic fibrils
by \cite{2007ApJ...655..624D}.
% The ranges of the physical quantities and the correlations
% in our simulations are in good agreement
% with the case with vertical magnetic field (Region 2)
% of Figs. 12 and 13 in their paper.
The properties of the spicules in quiet regions and coronal holes
were investigated by \cite{2012ApJ...750...16Z} and \cite{2012ApJ...759...18P}.
Both studies reported that the spicules became
longer in the coronal hole and shorter in the quiet region,
which was consistent with our results.
% The length of the spicule in these observations
% was roughly consistent with or
% slightly larger than that in our simulations.
We note that the parabolic trajectory of jets in our study
are consistent with the type I spicules
but inconsistent with the type II spicules,
which exhibit linear trajectory.
% We also note that the parabolic trajectory of jets in our study
% is not consistent with the type II spicules,
% which exhibit the linear trajectory.
Because the top of the spicule was considerably affected
by the observed wavelength
\citep{2014ApJ...792L..15P,2015ApJ...806..170S},
further studies are required
for the more quantitative comparison.
We also note that $T_c=0.4$ MK is too low for coronal holes
and $T_c=2$ MK is also too low for active regions.
The regional difference of density and
the spatial inhomogeneity of temperature and density
will explain at least a part of this discrepancy.
The magnetic field strength is also not
the typical value for these regions.
However, if the chromospheric jets are driven
by the acoustic waves from the lower atmosphere,
we expect that the magnetic field does not
affect the scale of the jets strongly.

% \subsection{理論研究との関係}

The numerical setting of the two-dimensional simulation
by \cite{2011ApJ...743..142H}
was very similar to that in our study.
The simulation with $T_\mathrm{c}=1$ MK
is consistent with Case B in their study.
% (Figure \ref{fig01}, panels (b), (e), and (h))
% is consistent with Case B in Figure 27
% of \cite{2011ApJ...743..142H},
% in terms of the ranges
% of the physical variables and correlations.
Their simulation employed
the Bifrost code \citep{2011A&A...531A.154G}
with more sophisticated treatment of radiation.
This agreement supports the validity of our simulations.
% \cite{2009ApJ...701.1569M} investigated the formation process
% of the chromospheric jets appearing
% in the three-dimensional simulation
% with the flux emergence from the convection zone.
\cite{2009ApJ...701.1569M} investigated
the chromospheric jets in three-dimensional domain
with the emerging magnetic flux.
The typical length of their jets
was approximately 1 Mm.
% These small jets were possibly produced
% by the high coronal temperature in their simulation.
% The typical length of their chromospheric jets
% was approximately 1 Mm.
The high coronal temperature produced by
the energy release of the flux emergence,
in addition to the rapid expansion
of the magnetic field lines in the corona,
is a possible explanation for why these jets were small.

% \subsection{重要だが考慮していない効果}

In this study, we neglect some physical processes
which are important to the chromosphere,
including non-equilibrium hydrogen ionization
\citep{2002ApJ...572..626C},
% formation of molecular hydrogen \citep{2011A&A...530A.124L},
% the effect of slippage between the neutral and ionized particles
% \citep{2012ApJ...750....6C,2012ApJ...753..161M},
more sophisticated treatments of radiation
\citep{2010A&A...517A..49H,2012A&A...539A..39C},
and three-dimensionality.
% Although these effects will change
% the results presented in this study quantitatively,
% we believe that the qualitative picture is not affected.
We believe that the qualitative picture is not affected.
The dependence on the grid size is investigated
and we find that the qualitative picture is not affected.
% The dependence on the grid size is also investigated
% by doubling the numerical resolution,
% and we find the similar dependence on the coronal temperature
% of the chromospheric jet motion.
% The width of the chromospheric jets are strongly affected
% by the numerical resolution,
% and further studies on the width are required.

\acknowledgments

This work was supported by JSPS KAKENHI Grant Number 15H03640,
``Joint Usage/Research Center for
Interdisciplinary Large-scale Information Infrastructures'',
``High Performance Computing Infrastructure'',
and the Program for Leading Graduate Schools, MEXT, in Japan.
Numerical computations were carried out
on the Cray XC30 supercomputer at Center
for Computational Astrophysics, National Astronomical Observatory of Japan.
The page charge of this paper is supported by Center for
Computational Astrophysics, National Astronomical Observatory of Japan.

% \appendix
% \section{Appendix material}




% \bibliographystyle{apj}
% \bibliography{apj-jour,reference}

\clearpage

%% the file, uncomment the \plotone, \plottwo, and
%% If you need a layout that cannot be achieved with \plotone or
%% \plottwo, you can invoke the graphicx package directly with the
%% \includegraphics command or use \plotfiddle. For more information,
%% please see the tutorial on "Using Electronic Art with AASTeX" in the
%% documentation section at the AASTeX Web site, http://aastex.aas.org/
%%
%% The examples below also include sample markup for submission of
%% supplemental electronic materials. As always, be sure to check
%% the instructions to authors for the journal you are submitting to
%% for specific submissions guidelines as they vary from

%% This example uses \plotone to include an EPS file scaled to
%% 80% of its natural size with \epsscale. Its caption
%% has been written to indicate that additional figure parts will be
%% available in the electronic journal.

\clearpage

\clearpage

\clearpage

% \clearpage

%% The following command ends your manuscript. LaTeX will ignore any text

\end{document}